\documentclass[aps,prl,reprint,showpacs,superscriptaddress]{revtex4-1}
\usepackage{amssymb}
\usepackage{amsmath}
\usepackage{graphicx}
\usepackage{textcomp}
\usepackage{xspace}
\usepackage{siunitx}
\usepackage[version=4]{mhchem}
\PassOptionsToPackage{usenames,svgnames}{xcolor}
\usepackage{xcolor}
\usepackage{pifont}
\usepackage{etoolbox}\usepackage{booktabs}
\newcommand{\ket}[1]{\left\vert#1\right\rangle}
\newcommand{\proj}[1]{\left\vert#1\middle\rangle\middle\langle#1\right\vert}
\newcommand{\tsub}[1]{\textsubscript{#1}}

\newcommand{\pMW}{\varphi_\text{MW}}
\newcommand{\pLO}{\varphi_\text{LO}}
\newcommand{\iu}{i}

\newcommand{\mS}{\mathrm{S}}
\newcommand{\mTo}{\mathrm{T_{0}}}
\newcommand{\mTp}{\mathrm{T_{+}}}
\newcommand{\mTm}{\mathrm{T_{-}}}

\newcommand{\mRp}{\mathrm{R_{+}}}

\newcommand{\To}{T\tsub{0}\xspace}

\newcommand{\msS}{\ket{\mS}}
\newcommand{\msTo}{\ket{\mTo}}
\newcommand{\msTp}{\ket{\mTp}}
\newcommand{\msTm}{\ket{\mTm}}

\newcommand{\msRp}{\ket{\mRp}}

\newcommand{\sS}{$\msS$\xspace}
\newcommand{\sTo}{$\msTo$\xspace}
\newcommand{\sTp}{$\msTp$\xspace}
\newcommand{\sTm}{$\msTm$\xspace}

\newcommand{\sRp}{$\msRp$\xspace}

\newcommand\drive[2]{$\ket{#1}\leftrightarrow\ket{#2}$}
\newcommand\driveS[2]{\drive{\csname m#1\endcsname}{\csname m#2\endcsname}}

\newcommand\todo[2][]{{	\color{blue}#2	\if\relax\detokenize{#1}\relax		\relax	\else		\footnote{\color{blue}#1}	\fi}}\newcommand\supplementary[2][]{~\footnote{See Supplemental Material at [URL] for #2}}

\graphicspath{{./}{./raw-figures/}}

\DeclareSIUnit\T{\tesla}
\DeclareSIUnit\ueV{\micro\eV}
\DeclareSIUnit\nW{\nano\watt}
\begin{document}
\title{Deterministic entanglement between a propagating photon
	and a singlet--triplet qubit in an optically active quantum dot molecule}\author{Y. L. Delley}
\affiliation{Institute for Quantum Electronics, ETH Z\"urich, CH-8093 Zurich, Switzerland.}
\author{M. Kroner}
\affiliation{Institute for Quantum Electronics, ETH Z\"urich, CH-8093 Zurich, Switzerland.}
\author{S. Faelt}
\affiliation{Institute for Quantum Electronics, ETH Z\"urich, CH-8093 Zurich, Switzerland.}
\affiliation{Laboratory for Solid State Physics, ETH Z\"urich, CH-8093 Zurich, Switzerland.}
\author{W. Wegscheider}
\affiliation{Laboratory for Solid State Physics, ETH Z\"urich, CH-8093 Zurich, Switzerland.}
\author{A. \.Imamo\u{g}lu}
\affiliation{Institute for Quantum Electronics, ETH Z\"urich, CH-8093 Zurich, Switzerland.}
\date{\today }

\begin{abstract}
Two-electron charged self-assembled quantum dot molecules exhibit a
decoherence-avoiding singlet-triplet qubit subspace and an efficient
spin-photon interface.
We demonstrate quantum entanglement between emitted photons and the spin-qubit after the emission event.
We measure the overlap with a fully entangled state to be \SI{69.5\pm2.7}{\percent},
exceeding the threshold of \SI{50}{\percent} required to prove the non-separability
of the density matrix of the system.
The photonic qubit is encoded in two photon states with an energy difference
larger than the timing resolution of existing detectors.
We devise a novel heterodyne detection method,
enabling projective measurements of such photonic color qubits along any direction on the Bloch sphere.
\end{abstract}
\maketitle
Spins in optically active quantum dots (QD) exhibit relatively short
$T_2^*$ coherence times. Despite this strong limitation, QDs stand
out among solid-state qubit systems for their excellent optical
properties that render them promising for quantum communication
tasks relying on a quantum interface between stationary (spin) and
flying (photonic) qubits.
Recent experiments have used this favorable feature to demonstrate
coherent all-optical spin manipulation~\cite{Press-NP-2010},
emission of entangled photon-pairs~\cite{Young-JoAP-2007,Hafenbrak-NJP-2007},   spin-photon entanglement~\cite{Gao-N-2012,DeGreve-N-2012,Schaibley-PRL-2013},
teleportation from a photonic to a spin qubit~\cite{Gao-NC-2013}
and heralded distant spin entanglement using QDs in Voigt geometry~\cite{Delteil-NP-2016}.
However, the magnetic field configuration to achieve efficient spin measurement~\cite{Delteil-PRL-2014}
is incompatible with the configuration for coherent manipulation.

The moleclular states \sS and \sTo in optically
active quantum dot molecules (QDMs) in Faraday geometry emerge as a promising
alternative effective qubit for quantum information processing since (i) they exhibit
a decoherence-avoiding clock-transition that is insensitive to
fluctuations in both electric and magnetic fields~\cite{Weiss-PRL-2012},
(ii) the spin polarized triplet states (\sTp and \sTm) of the
ground-state manifold exhibit cycling optical transitions~\cite{Delley-arXiv-2015}, and
(iii) the qubit states exhibit equal coupling strength to common optically excited
trion states, essential for maximal spin-photon entanglement.
In this letter, we experimentally determine the amount of entanglement obtained
from the spontaneous emission from such an excited state.

\begin{figure}
\begin{center}
\includegraphics[width=\columnwidth]{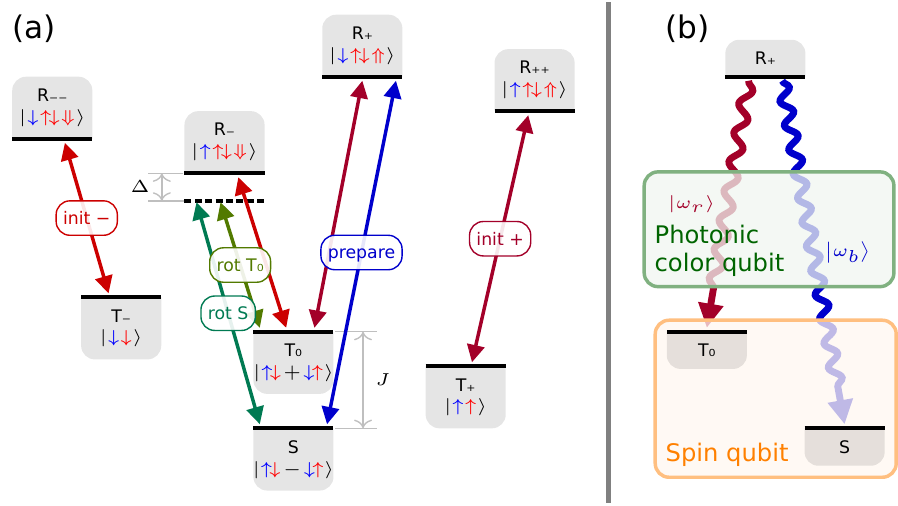}
\end{center}
\caption{
(a) Spin states and optically excited states in the (1,1) regime.
In the state kets, single arrows denote electron spins, double arrows denote hole spins.
Spins denoted by a red (blue) arrow are mainly located in the top (bottom) dot,
indicating that we optically excite only the top QD.
The transitions that are driven in this experiment are labelled with the laser that drives it.
\emph{init~+} is simultaneously resonant with \driveS{To}{Rp} and \driveS{Tp}{Rpp},
while \emph{init~−} is resonant with \driveS{To}{Rm} and \driveS{Tm}{Rmm}.
They are used to pump the spin into \sS, as well as to measure population in \sTo.
\emph{prepare} excites \sRp from \sS, as well as is used to measure population in \sS.
\emph{rot~Tₒ} and \emph{rot~S} perform coherent spin rotation,
and are red detuned by $\Delta$ from \driveS{To}{Rm} and \driveS{S}{Rm} respectively.
(b) The effective spin qubit is made up of the states \sS and \sTo.
Decay from the excited state \sRp projects the spin into an entangled state with
the photonic color qubit in the space spanned by $\ket{\omega_r}$ and $\ket{\omega_b}$.
}
\label{fig:QDM-levels}
\end{figure}

\paragraph{S--\To qubits in QDMs}
Our experiment is carried out on a single \ce{InGaAs} self-assembled QDM,
consisting of two QDs separated by a \SI{9}{\nm} \ce{GaAs} tunneling barrier
along the growth direction~\cite{Xie-PRL-1995}.   The QDM is embedded in a Schottky diode, formed by a semi-transparent metallic top gate
and an \textit{n}-doped layer,
which is used to control the charge state of the QDM
and the optical transition energies vie the quantum confined Stark effect~\cite{Miller-PRL-1984}. A distributed Bragg reflector (DBR) below the doped layer
forms a weak planar microcavity together with the top gate,
enhancing the collection efficiency though a combination of emission profile modification
and Purcell effect~\cite{Delteil-PRL-2014}.
Thanks to engineered confinement energies of the two QDs,
the QDM can be brought into the (1,1)-regime~\cite{Kim-NP-2011,Greilich-NP-2011},
where each QD is charged with a single electron.
In this regime, the singlet state (\sS) is split from the triplet states (\sTo, \sTp and \sTm) by the exchange splitting,
which is gate-voltage tunable and has a minimum value of $J = \SI{97}{\GHz}$ in our device.
The triplet states are split by $\SI{15.5}{\GHz}$ from each other by an external magnetic field of \SI{2}{\tesla} that is applied along the growth direction (Faraday geometry).
The relevant level scheme is outlined in fig.~\ref{fig:QDM-levels} (a).
Under these conditions, \sS and \sTo can be compared to atomic clock transitions that are insensitive to both electric and magnetic field fluctuations~\cite{Weiss-PRL-2012}.
Coupling to the common optically excited state \sRp (with equal oscillator strength) allows for coherent manipulation of the qubit~\cite{Kim-NP-2011,Greilich-NP-2011}.
Spontaneous radiative recombination of \sRp projects the joint-system of the QDM and the optical field into an entangled state
\begin{equation}
\ket{\Psi} = \frac{1}{\sqrt{2}}\left( \msS \ket{\omega_b} - \msTo \ket{\omega_r} \right),
\label{eq:entangled-state}
\end{equation}
where $\ket{\omega_b}$ and $\ket{\omega_r}$ denote single-photon states with center frequencies at $\omega_b$ and $\omega_r$ respectively,
both of them being circularly polarized with the same handedness.
The relative phase of the state is fixed by the optical selection rules.

\begin{figure}
	\includegraphics[width=\columnwidth]{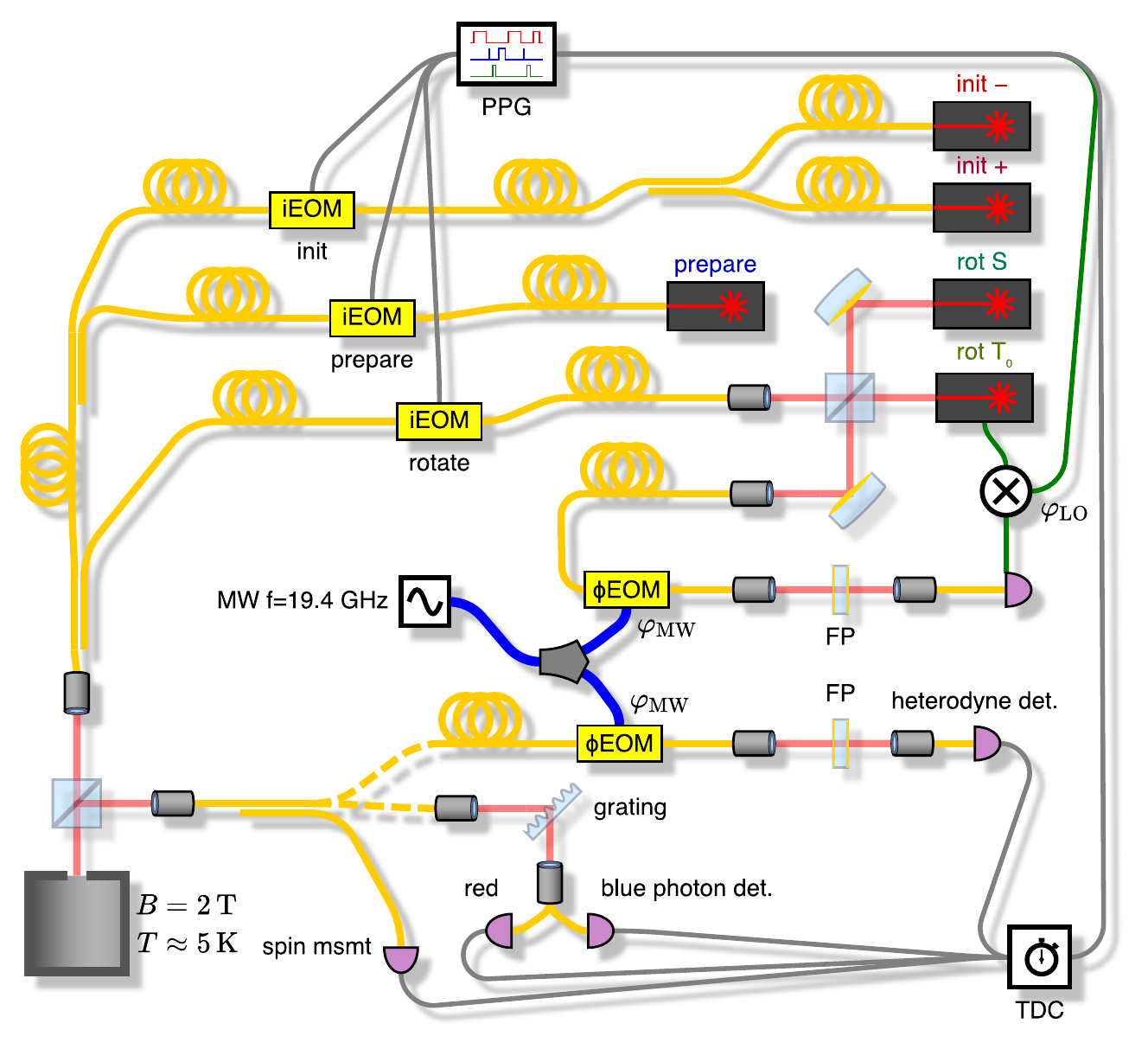}
	\caption{
		Schematic of the experimental setup. 						Abbreviations used are: \emph{PPG:} Pulse pattern generator,
		\emph{TDC:} time to digitial converter,
		\emph{iEOM:} electro-optic intensity modulator,
		\emph{ΦEOM:} electro-optic phase modulator,
		\emph{FP:} Fabry-Pérot etalon.
		Classical correlations and quantum correlations are measured sequentially,
		the fiber of the corresponding photon-detection setup is connected,
		the other is disconnected.	}
	\label{fig:setup}
\end{figure}

\paragraph{Experimental setup}
Figure~\ref{fig:setup} shows schematically how the experiment is set-up.
The sample is held in a liquid helium bath cryostat at about \SI{5}{\K}.
A confocal microscope with NA=\num{0.68} is used to direct \si{\ns}-pulses from tunable diode lasers 
to the QDM to manipulate the qubit state
and to read it out via detection of resonance fluorescence (RF).
The laser pulses are linearly polarized, and the RF is collected from the orthogonal linear polarization,
such that all transitions couple to excitation and detection equally well~\cite{Vamivakas-N-2010,Yilmaz-PRL-2010},
while reflected laser light is suppressed by a factor of \num{e6}.
Multiple \ce{Si} avalanche photo-diodes (APDs) are used
to detect the emitted photons,
and a time-digital converter (TDC) records the arrival time of every photon.

\paragraph{Entanglement generation and verification}
Resonant laser pulses are used to initialize the QDM into \sS via optical spin-pumping~\cite{Atature-S-2006}.
To this end, we combine the light of two lasers,
one of which is resonant with \driveS{To}{Rp} and \driveS{Tp}{Rpp},
the other is resonant with \driveS{To}{Rm} and \driveS{Tm}{Rmm}.

From \sS, a \SI{375}{\ps} long resonant laser pulse prepares the QDM in \sRp.
Spontaneous emission creates the entangled state between the propagating photon and
the remaining spin-qubit in the \sS--\sTo subspace.

To verify the entangled state,
we estimate the overlap of the post-emission state described by the density matrix $\rho$
with the entangled state $\ket{\Psi}$ as defined in equation~(\ref{eq:entangled-state}),
quantified by the state fidelity
\begin{equation}\begin{split}
F &= \text{Tr}(\rho\proj{\Psi}) \\
  &= \frac{1}{2} \underbrace{\left(
	\rho_{\mS b,\mS b}+\rho_{\mTo r,\mTo r}
	\right)}_{F_\|} \\
  &+ \frac{1}{2}\underbrace{\left(
	\rho_{\mS b,\mTo r}+\rho_{\mTo r,\mS b}
	\right)}_{F_\bot}.\label{eq:fidelity-def}
\end{split}\end{equation}
The fidelity can be decomposed into two parts:
$F_\|$ quantifies the amount of \emph{classical} correlations between the probabilities
of finding each subsystem in a particular eigenstate.
$F_\bot$ on the other hand is sensitive to the relative phase between the two parts
of the state $\ket{\Psi}$ when written in the eigenbasis,
thus we refer to that term as the \emph{quantum} correlations.
The state described by $\rho$ generally is a mixture of pure states,
such that $F$ becomes a convex mixture of the fidelity of each contained pure state.
Since no separable state has a higher fidelity than $1/2$,
observing a fidelity above that limit proves inseparability of $\rho$~\cite{Sackett-N-2000}.  
\paragraph{Measurement of the classical correlations}
To estimate $F_\|$, we measure the correlations between the two subsystems in their eigenbases.
To that end, we disperse the photon from the entanglement pulse using a transmission grating,
split the two color components using a fiber-bundle with two cores next to each other,
and detect it using a dedicated APD for each component.

To measure the spin, we detect the scattered photons from spin-selective RF~\cite{Vamivakas-NP-2009},
where either a laser pulse resonant with \driveS{S}{Rp} is used to detect the state \sS,
or a combined two-laser pulse resonant with \driveS{To}{Rp} and \driveS{To}{Rm} is used to detect the state \sTo.
Contrary to spin initialization, for spin measurement, any one of the two lasers would suffice.
However, that would have required an additional EOM and a pulse-pattern channel, neither of which were easily available.

\begin{figure}
	\begin{center}
		\includegraphics[width=\linewidth]{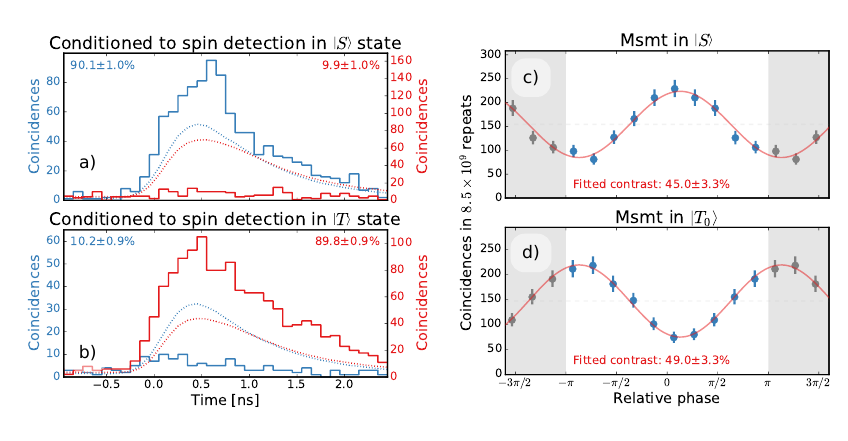}
	\end{center}
	\caption{
		\emph{(a):} Solid lines: Histogram of coincidences between spin detection in \sS state and photon
		color detected to be $\ket{\omega_r}$ (red line) and $\ket{\omega_b}$ (blue line).
		Dotted lines: Expected coincidence counts from individual detection rates,
		disregarding spin measurement.
		The two photon colors have separate vertical scales to account for the different sensitivities
		of the two detection paths.
		\emph{(b):} As \emph{(a)}, but spin detection in \sTo state.
		Both datasets were measured simultaneously,
		and are histogrammed from \num{2.7e10} repetitions each,
		acquired over a time period of about \SI{12}{\hour}.
		\emph{(c) and (d):} Dots and bars: Total number of coincidence events between the beat-phase sensitive
		photon detector and post-rotation measured spin in \sS and \sTo respectively,
		plotted against the relative phase between the photon detection and the spin rotation pulse.
		Vertical lines denote \SI{68}{\percent} confidence intervals derived from Poissonian statistics.
		Shaded areas outside the interval $[-\pi,\pi]$ contain replicas of the data points displayed inside this interval.
		Solid lines: Maximum-likelihood fit of a sinusoid to the data.
		Both datasets were simultaneously fitted with a fixed relative phase shift of $\pi$,
		but individual mean and amplitude.
	}
	\label{fig:Zbasis-results}
	\label{fig:Xbasis-results}
\end{figure}

Figure~\ref{fig:Zbasis-results} (a) and (b)  show the time resolved fluorescence measured by the spectrally filtered detectors,
conditioned on the observation of a photon during the following spin measurement pulse.
We normalized the values using the relative overall sensitivity of the two detection paths,
which we determined separately\supplementary{Spectral selectivity of the grating-filtered photon detection}.
There is a strong correlation between the spin and the detected photon color:
$P(b|\mS)\mathbin{:}P(r|\mS) = (\num{90.1}\mathbin{:}\num{9.9})\SI{\pm1.0}{\percent}$ and
$P(b|\mTo)\mathbin{:}P(r|\mTo) = (\num{10.2}\mathbin{:}\num{89.8})\SI{\pm0.9}{\percent}$.
Errors are one standard deviation and are derived from counting statistics.
We take $P(b)+P(r)=1$ as well as $P(S)+P(T)=1$,
valid if the results are conditioned to the cases where a photon is detected,
and as long as optically forbidden transitions are rare.
From this, we can extract a lower bound to the fidelity as
$F_\| = \left[P(b|\mS)P(\mS)+P(r|\mTo)P(\mTo)\right] \geq \min\left\lbrace P(b|\mS),P(r|\mTo) \right\rbrace=\SI{89.4\pm0.8}{\percent}$
since $F_\|$ is a convex mixture of the two conditional probabilities.

We attribute the reduction from perfect correlations mainly to double excitation during the entanglement pulse:
The duration of the excitation pulse is comparable with the exciton lifetime of about \SI{400}{\ps},
such that emission events early during the pulse may loose the correlation with the spin during subsequent excitation events.
Simulation of the optical Bloch equations suggest between \SIrange{5}{10}{\percent} double-excitation events.

\paragraph{Measurement of quantum correlations}
To determine $F_\bot$, we measure both the photon and the spin in a superposition basis,
where the basis states are lying on the equator of Bloch spheres of the subsystems.
Entanglement between the systems then implies a sinusoidal dependency of coincidence probabilities
on the relative azimuthal angle between each system's measurement basis~\cite{Blinov-N-2004}.

To measure the photon in a superposition state of the two colors,
we need to detect the phase of the beat-note at $\omega_J = 2\pi\cdot \SI{97}{\GHz}$.
This is faster than the timing resolution of existing photo-detectors.
We therefore employ a heterodyne detection scheme:
Using an electro-optic phase modulator (phase-EOM) driven by a microwave (MW) signal,
we generate sidebands to each spectral component of the incoming photon.
The MW frequency $\omega_\text{MW}$ is close to $\omega_J$ divided by an integer;
due to limitations of our MW source, we chose $\omega_J/5$.
A free-space Fabry-Pérot etalon with a free spectral range
of \SI{200}{\GHz} and a bandwidth of \SI{5}{\GHz} allows us to single-out
the $k=+3$ sideband of the red photon simultaneously with
the $k=-2$ sideband of the blue photon.
Hence, detection of a photon after the etalon corresponds to
a projective measurement of the incoming photon in a superposition state
whose phase is given by $5\pMW$, determined only by the MW source.
In the rotating frame of the two-color photon, this phase rotates
at a frequency of $5 \omega_\text{MW} - \omega_J \approx 0$,
which can be chosen arbitrarily.
It was set to \SI{131}{\MHz},
such that the timing jitter of standard \ce{Si}-APDs
does not affect the measured visibility.

To measure the spin in a superposition state,
we rotate the spin by $\pi/2$
around a vector in the equatorial plane of the Bloch sphere,
and then measure the population in the \sS or \sTo states via RF detection.
To that end, we utilize a variation of the standard method of ultrafast coherent optical control
based on the optical (AC-)Stark effect (see \cite{Press-N-2008} and references therein):
Instead of using a single \si{ps} pulse that is far-detuned from both transitions,
we use two quasi-resonant \si{ns} pulses detuned by $\Delta \approx \SI{10}{\GHz}$ from the transitions
\driveS{S}{Rm} and \driveS{To}{Rm} to induce an energy shift to a coherent superposition of the states \sS and \sTo.
In this case, the azimuthal angle of the rotation vector is determined by the relative optical phase of the two laser pulses,
which stays constant during the pulse in the rotating frame of the spin.
To ensure a fixed phase relation over the whole measurement time,
we embed one of the two diode lasers in a phase-locked loop (PLL),
relying on the same heterodyne detection method employed for the photon measurement:
The \SI{97}{\GHz} beat-note of the two lasers is down-mixed via side-band generation and spectral filtering,
where the MW signal is derived from the same source that drives the photon measurement.
The resulting beat-note of \SI{131}{\MHz} is then locked to a local oscillator (LO) that is synchronized with the pulse sequence.
While the phase of the MW source $\pMW$ therefore determines the azimuthal angle of both the photon measurement basis as well as the spin measurement basis,
the relative angle between the two is only determined by the LO phase $\pLO$.

With this detection scheme, given an arbitrary joint density matrix $\rho$,
the probability to detect a coincidence between the photon measurement and the spin measurement is given by
\begin{equation*} \begin{split}
P_\text{coinc} &\propto 1  + \mathcal{R}\biggl[ e^{\iu \pLO} \rho_{\mS b,\mTo r}   + e^{2 \iu \pMW} \rho_{\mS r,\mTo b} \\
  &+ e^{\iu \pMW} \left(
	  \rho_{\mS b,\mS r}
	  + \rho_{\mTo b,\mTo r}
  \right) \\
  &+ e^{\iu \left(\pMW + \pLO\right)} \left(
	  \rho_{\mS r,\mTo r}
	  + \rho_{\mS b,\mTo b}
  \right)   \biggr]. \end{split} \end{equation*}
By letting the MW source run freely, we average over $\pMW$,
such that the modulation depth of the coincidence rate with respect to $\pLO$ is a direct measure of $F_\bot$.
Figure~\ref{fig:Xbasis-results} (c) and (d) shows the detected photon counts conditioned on detection
of a scattered photon during the spin measurement,
depending on the phase $\pLO$.
Fitting the data by a sinusoidal function,
we extract a visibility of \SI{45.0\pm3.3}{\percent} when the spin is measured in the \sS state,
and \SI{49.0\pm3.3}{\percent} when the spin is measured in the \sTo state.
Errors are one standard deviation and are derived from counting statistics.
Since the measurement process can only make the visibility contrast worse,
the higher value of the two puts a lower bound on
$F_\bot\geq \max \left\lbrace v_{\mS}, v_{\mTo} \right\rbrace = \SI{49.5\pm2.9}{\percent}$.

This value is considerably lower than the ideal case for perfect entanglement.
A significant amount of visibility is lost due to the spectral filtering properties
of the Fabry-Pérot etalon used in the phase-sensitive photon detection setup,
limiting the visibility of the $F_\perp$ measurement to about \SI{84}{\percent}:
The beating of neighboring pairs of sidebands are out of phase by $\pi$.
Our FP etalon only suppresses the neighboring pairs by a factor of 12\supplementary{Spectral selectivity of the Fabry-Pérot etalon}.
The limitation can be overcome by using higher finesse FPs or higher MW frequencies. 
A further reduction of the visibility is likely to be due to imperfections of the spin-rotation pulse.
(See supplementary material for a detailed listing of all sources of error\supplementary{Estimation of imperfections}).

Combining the two measurements,
we obtain $F=\SI{69.5\pm2.7}{\percent}$,
where the error indicates one standard deviation of uncertainty due to counting statistics.

To summarize, we have shown deterministic generation of a photonic color qubit entangled with
the QDM spin qubit.
Working with color qubits split by a large energy separation was enabled thanks
to our heterodyne quadrature detection method, effectively erasing the energy separation.
Since the \sS--\sTo spin qubit can controllably be furnished with a dipole,
QDM's promise to bridge the gap between optical long-distance quantum communication
and quantum information processing in the solid state.
Possible candidates for coupling to the QDM dipole are
(a) the dipole of quantum-well exciton-(di-)polaritons in planar microcavities~\cite{Christmann-APL-2011},
(b) the electric field of photons in microwave cavities~\cite{Tsuchimoto-2017} and
(c) phonons of microresonators via the piezoelectricity of \ce{GaAs}.\begin{acknowledgments}We acknowledge helpful discussions with Joseph M. Renes and Emre Togan.
This work is supported by NCCR Quantum Science and Technology (NCCR QSIT),
research instrument of the Swiss National Science Foundation (SNSF) and
by Swiss NSF under Grant No. 200021-140818.\end{acknowledgments}
\makeatletter
  \close@column@grid
  \cleardoublepage
  \twocolumngrid
\makeatother	\section{Supplementary Information}
\subsection{Spectral selectivity of the grating-filtered photon detection}

To measure the classical correlations, the color of the photon has to be resolved.
This has been achieved using spectral filtering with the help of a transmission grating.
To that end, the collected light, after dividing using a 50:50 fiber-splitter
for spin detection, is collimated into a free-space beam of \SI{\sim10}{\mm} diameter.
The first order diffraction generated by the transmission grating is focused using a \SI{250}{\mm} focal length
achromatic doublet onto the facet of a multi-mode fiber bundle with two cores
separated by \SI{105}{\um}.

\begin{figure}
	\begin{center}
		\includegraphics[width=\linewidth]{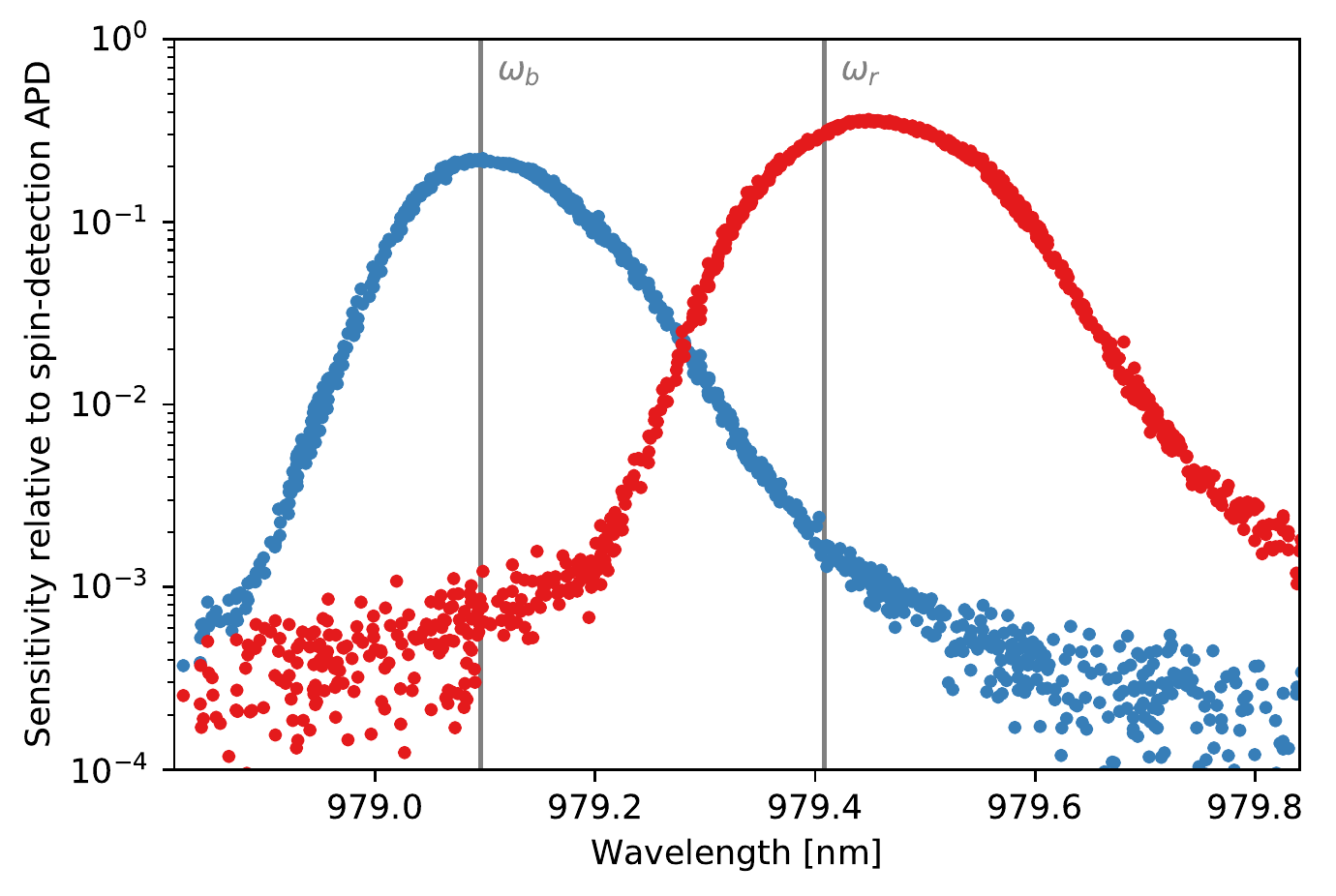}
	\end{center}
	\caption{
		Spectral sensitivity of photon detectors used in measurement of classical correlations,
		relative to the sensitivity of the detector used for spin measurement.
	}\label{fig:grating-spectrum}
\end{figure}

The sensitivity of that setup can be calibrated by comparing the count-rate
of the APDs connected to the multi-mode fibers with the count-rate of the APD used for spin measurement,
while laser light of a known wavelength is coupled into the setup.
Figure~\ref{fig:grating-spectrum} displays a full spectrum obtained using this method.
It indicates of a suppression of more than two orders of magnitude.
Thus, the resulting probability for a wrong color assignment is small enough
to not significantly impact the measurement of the classical correlations.
As the response is sensitive to the alignment of the free-space elements,
the ratio between the sensitivities for detection of a photon at $\omega_r$ and $\omega_b$
was monitored during the measurement via the magnitude of the residual laser-background
during the spin-measurement pulses.
It was confirmed to stay constant within the measurement accuracy of \SI{\pm 1}{\percent}.

\subsection{Spectral selectivity of the Fabry-Pérot etalon}\label{sec:FP-sensitivity}

\begin{figure}
	\begin{center}
		\includegraphics[width=\linewidth]{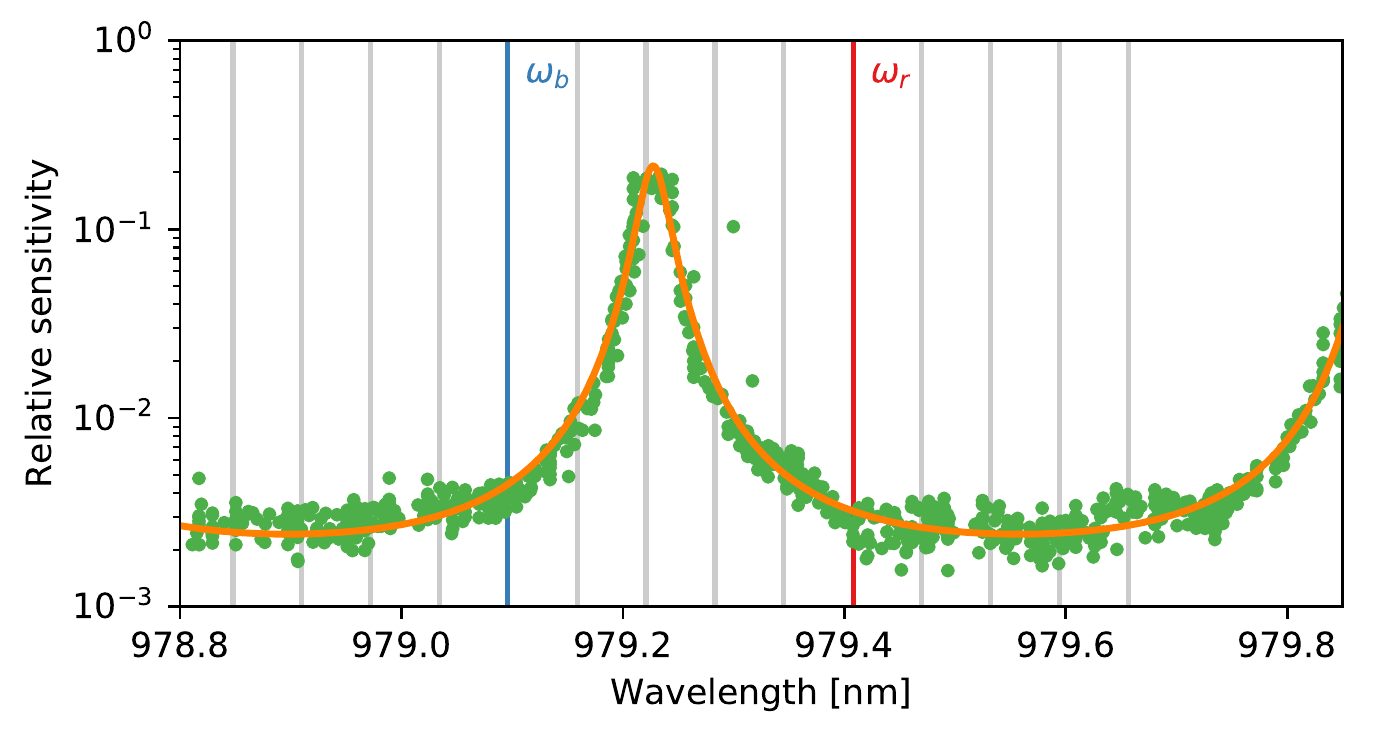}
	\end{center}
	\caption{
		Spectral sensitivity of the detector
		for the superposition-basis photon measurement,
		relative to the sensitivity of the detector used for spin measurement.
		The solid orange line shows a fit based on the transmission model of a \SI{200}{\GHz} FSR
		Fabry-Pérot etalon with a finesse of 200.
		The vertical lines indicate the wavelength of the two color components
		of the entangled photon, along with the sidebands generated by the phase EOM.
	}\label{fig:FP-spectrum}
\end{figure}

The spectral response of the Fabry-Pérot filter used in the measurement of the quantum correlations
has been measured using using a tunable narrow band laser.
It is displayed in figure~\ref{fig:FP-spectrum}.
Here, the suppression of the next pair of sidebands separated by \SI{19.4}{\GHz} is
only \num{\sim12},
leading to a reduction of the visibility of the quantum correlations of the order of \SI{16}{\percent}.

\subsection{Phase stability of the setup}

The measurement of the quantum correlations can be considered
an interferometric measurement in the microwave domain.
In this view, the goal of the experiment is to measure the stability
of the relative phase between the emitted photon and the QDM spin.
The phase reference for the spin is given from the phase of the beat-note
of the rotation pulse, which is locked to the MW phase (up to a tunable phase given by the LO),
which in turn determines the phase of the photon measurement.
The precision up to which the phase relation between spin and photon
can be measured is determined by the phase stability of the interferometer loop,
which includes many elements and is of the order of \SI{30}{\m} long.

We measured the stability of the whole interferometer
by using the photon-superposition detection setup
to detect light from the rotation lasers, reflected by the sample surface.
The photon detection rate should vary sinusoidally with the LO phase.
The visibility of the modulation allows to determine a lower bound on the short-term stability of the loop,
while slow drifts in the phase offset can directly be observed.
The QDM was moved out of the focus to prevent any RF interfering with the measurement.
Over the course of 12 hours, the visibility stayed constant at \SI{82}{\percent},
but the phase offset drifted by \SI{12}{\degree}.

The limited visibility can have many reasons,
among them, the spectral selectivity of the Fabry-Pérot etalon
or an imbalance between the powers of the two lasers participating in the rotation pulse.
Thus, the visibility is not an independent measurement of the interferometer stability.
On the other hand, the phase drift only depends on the stability of the interferometer,
but the loss of visibility due to averaging over \SI{12}{\degree} is significantly below one percent.

The phase stability of the laser phase lock at short time-scales has been measured separately
by comparison of time-traces of the two inputs of the feedback controller;
the beat note as measured by the photo diode and the LO reference.
We determined an RMS error of \SI{8}{\degree} with a correlation time of about \SI{1}{\us}.
This amounts to a reduction of the visibility by approximately \SI{1}{\percent}.

\subsection{Spin-rotation fidelity and spin coherence}

The spin-rotation pulse has a duration of \SI{2.5}{\ns}
and was placed with a delay of \SI{3}{\ns} to the entanglement pulse,
in order to prevent residual laser photons impacting the photon measurement.
Due to limited $T_2^*$ coherence,
both the pulse duration and the delay time lead to a decrease of the quantum correlations.
In a separate Ramsey experiment, we determined a coherence time of \SI{9}{\ns},
with an exponential decay of fringe visibility.
Hence, dephasing during the rotation pulse limits the visibility to \SI{87}{\percent},
while dephasing during the delay reduces the visibility to \SI{71}{\percent},
leading to an overall visibility of \SI{62}{\percent}.

The spin coherence could be prolonged significantly by employing spin-echo.
We measured several hundreds of \si{\ns} coherence time on this QDM
under spin-echo.
However, with the parameters of our rotation pulses,
that would require a \SI{5}{\ns} pulse,
whose fidelity would again be impacted by the $T_2^*$ time.
In principle, our quasi-resonant rotation pulses could be shortened down to a few hundred \si{\ps},
limited by the EOM bandwidth.
To keep the pulse-area constant, the pulse power would have to be increased,
which was not possible in our case.
Instead, ultrafast spin rotation could be employed~\cite{Press-N-2008},
but in this case, the large exchange splitting requires even higher laser-powers,
which proved to be problematic as well.

\subsection{Estimation of imperfections}

Arguably, all of the limitations described above and summarized in table~\ref{tab:imperfections}
are of technical nature that can be overcome using improved methods and equipment.
Together, they account for nearly all of the reduced visibility of quantum coherence in our measurements,
suggesting that the intrinsic entanglement fidelity is close to unity.

\begin{table}
	\begin{center}
	\begin{tabular}{p{0.55\linewidth} S}
		Imperfection & \si{\percent} \\ \toprule
		Spin dephasing 		& 29 \\
		Rotation pulse 		& 13 \\
		Sideband suppression
			    & 16 \\
		Phase lock performance & 1 \\
		Phase reference drift& 0.2 \\ \midrule
		Remaining visibility & 52 \\
	\end{tabular}
	\end{center}
	\caption{Estimation of impact on measured quantum correlation visibility of various imperfections}
	\label{tab:imperfections}
\end{table}

\end{document}